\documentclass[10pt]{article}

\usepackage{cite} 

\usepackage{graphicx}
\usepackage[font=scriptsize]{caption}

\usepackage[a4paper,top=2cm,bottom=2cm,left=3cm,right=3cm]{geometry} 

\usepackage{hyperref}

\begin{document}

\title{\textbf{Introducing General Relativity in high school: a guide for teachers}\vspace{-4ex}}
\date{}
\maketitle
\begin{center}
\author{Adriana Postiglione$^{1,2}$, Ilaria De Angelis$^{1,2}$}
\end{center}

\paragraph{} \parbox[t]{1\columnwidth}{$^1$Dipartimento di Matematica e Fisica, Universit\`a degli Studi Roma \\Tre, Rome (Italy)\\%
    $^2$INFN Sezione di Roma Tre, Rome, (Italy)\\
    
    adriana.postiglione@uniroma3.it}

\begin{abstract}
Introducing Modern Physics represents an increasingly urgent need, towards which physics education concentrates many efforts. In order to contribute to this attempt, at the Department of Mathematics and Physics of Roma Tre University in Rome we focused on the possibility of treating General Relativity (GR) at high school level. We started with an interactive activity addressed to students that exploits the rubber sheet analogy (RSA) to show various phenomena related to gravity using the concept of space-time. Then, having verified its effectiveness, we began to include it among the initiatives the Department carry for high school teacher professional development, with the explicit aim of making them capable of carrying on the activity autonomously in the classrooms. In this paper, we analyse the teacher training approach we realized, and all the materials developed.

\end{abstract}


\section{Introduction}
Among the various actions physics education research is carrying on in recent years, introducing Modern Physics in high school certainly stands out. The urgency of this effort is evident also in Italy, where the Ministry of Education itself added Modern Physics topics in the program directions provided to teachers starting from 2010 \cite{Indicazioni}.
In order to contribute to this collective effort, at the Department of Mathematics and Physics of Roma Tre University, we started to work on the possibility of talking about General Relativity (GR) with a non-expert public. We began with groups of people of various ages (adults, teenagers, children), we then involved groups of high school students and we finally turned to high school teachers, with the ambitious aim of making them confident and autonomous in treating GR in their classrooms. At the end of our study, we obtained a training method that provides teachers with the right tools and guidelines, which already proved to be well received by them. In this paper we describe the training method we developed and the steps that led us to build it.
The model we used is the popular rubber sheet analogy (RSA), that compares the Einsteinian space-time to a rubber sheet that stretches under the weight of a mass. Using marbles and balls thrown on the sheet, it is possible to simulate the gravitational attraction of several astrophysical objects. This analogy has been studied for years in physics education research, proving to have weaknesses and critical points \cite{Freefall, Curved, Price, Gould} but also to be effective and potent in many cases \cite{Possel, Farr, Baldy, Thorne, Postiglione1}.  
For this reason, at the Department we decided to build a 1.8 m diameter circular structure supporting a lycra sheet that could show the RSA. Then, we started to use it with marbles and balls of different weights, with the aim of reproducing various phenomena related to gravity: from Kepler's laws to space exploration, from gravitational lensing to black holes.  Our intention was to build an interactive activity that would allow the audience to first-hand 
experience the phenomena we wanted to deal with, since this kind of activity proved to be effective and engaging in learning \cite{Freeman,Prince,Hake}. 

Once we had an idea of what we could do with the structure, we started testing it with a varied audience, initiating our preliminary study, which eventually led us to develop a real teachers’ professional development initiative.
The remaining paper is organized as follows. In section 2 we retrace the preliminary study we conducted, focussing on different audiences. In section 3 we present the approach we tested in presence with teachers; in section 4 we illustrate some first reactions to our training activity and in section 5 we present our conclusion and some future developments of our work.

\section{The preliminary study}
Since our final aim consisted in structuring a teachers’ professional development course involving the use of the RSA, we needed to know as much as possible about the reactions of the audience to it: their understanding, feedback, degree of involvement, the possible birth of misconceptions and the best way to solve them. For this reason, we started our study with a varied audience, before turning to high school students and teachers.

\subsection{Experimentation with a varied audience}
Our test with a varied audience (adults, teenagers, children) was conducted during the public outreach events that take place at the Department three times a year. During these events the Department opens its doors to the public and University students, researchers and professors show different experiments and scientific activities. The atmosphere is typically very informal, and participants can move freely between the proposed activities, asking questions and interacting with scientists. 
In this context, the activity we carried out with the rubber sheet dealt with a variety of laws and phenomena (the motion of planets around stars, binary systems, black holes, gravitational lensing, time dilation) and was very well received: people of different ages (from 3 years old children to adults) were intrigued by the structure and tried to arrange weights on the sheet and throw marbles in order to see their behaviour. Moreover, several people asked questions also related to complex topics such as black holes or gravitational lensing. 
From these tests we thus concluded that the playful and interactive aspect of our activity represents its most relevant strength, since it attracts and intrigues people, who are then willing to listen to further insights provided through videos and images. Moreover, we noticed that a significant part of the public's curiosity came from discovering the orbit of planets or space probes rather than just the exotic topics like black holes. This means that the rubber sheet could also be used effectively to talk about classical gravity and not just GR.

\subsection{Experimentation with high school students}
The experimentation with high school students involved 14 Italian high school different classes and 2 mixed groups of students coming from different high schools, for a total of about 400 students.
We initially used our rubber sheet with the participants of a physics summer school. They were selected taking into account their interest in physics and their previous grades, and they all already attended the third year of Italian high school (meaning that they already treated the Newtonian description of gravitational force with their teacher). After a short activity in which we explained to them the basics of the RSA, the space-time structure was placed in a dedicated room that we could easily access with them. In the following days, we often returned to the space-time simulator, asking participants to experience different phenomena and try to explain them. In this way we were able to closely follow the students' understanding of the topics, and we were able to calmly reason about their misconceptions and the way to overcome them.
After the experience with the summer school, we decided to define a more structured activity lasting an hour and a half that could deal with several topics related to gravity: Kepler's laws, gravity assist, binary systems, gravitational lensing, black holes. In order to do that, we decided to validate it with classes of different years (one class for each of the five years of the Italian high school) thanks to the collaboration of a teacher who has been working with the Department for years. In this way we investigated the reaction of participants of different ages.
Once we had structured the activity, we began to verify its effectiveness by involving other schools, reaching more than 200 students of 9 classes. Before and after the activity we administered questionnaires to the participants, in order to explore their reactions and comprehension of the different topics addressed. As it will be clear in the following of this paper, the results of such tests were precious in order to build an effective teachers’ professional development activity, so we will report their essential aspects; the complete analysis of the questionnaires can be found in \cite{Postiglione1}.

\subsection{Experimentation with high school teachers }
In addition to discussing with the teachers who accompanied their classes during the experimentation with students, in the same period of the tests described so far, we carried on a parallel experimentation with high school teachers. In particular, we involved the participants to the teachers’ professional development course that was taking place at the Department, and that was focused on Modern Physics. The participants were about 25 and came from schools of Roman area. During the course, we showed the teachers our space-time simulator and the activity related, in order to train them on GR. The reactions were very positive: the teachers demonstrated to be curious and interested; some of them also claimed that the RSA provides a wonderful way to finally visualize Kepler's laws. When asked explicitly, they replied that they would be very interested in including the activity in their teaching.

\subsection{Results from the preliminary study}
The experimentation with different audiences convinced us that our activity based on the use of the rubber sheet could really be useful for teachers to deal with gravity in their classrooms. In fact:
\begin{enumerate}
    \item Our use of the RSA was effective in talking about gravity, as demonstrated by the results of the questionnaires administered to high school students \cite{Postiglione1}. The phenomena treated were visualized better, understood and remembered longer compared to a traditional treatment.

    \item The activity proved to engage and entertain the participants, who were thus more willing to ask questions and interact.

    \item High school teachers demonstrated to be interested in including the activity in their teaching.

    \item Although the RSA could reinforce misconception and conceptual mistakes related to gravity, they could be also easily overcome if addressed explicitly. Therefore, teachers had to be prepared to know and react to these aspects.

\end{enumerate}

\section{The teachers’ professional development course}
In structuring the teachers’ professional development course, an important role was played by the results of the questionnaires we administered to the high school students who participated in our activity with the RSA. In particular, we administered three questionnaires: one before the activity, one immediately after and one four months after, so that we could investigate students’ understanding and follow its evolution over time. Out of about 9 classes who participated (for a total of about 200 students) we collected 154 answers. While the complete analysis can be found in \cite{Postiglione1}, here we summarize the results relevant to the aim of this paper.
The first element that is evident from the questionnaires is the fact that the RSA approach helps to comprehend Newtonian gravity, already typically treated in school. This is particularly true concerning Kepler's second law, that was visualized better through the motion of the marbles on the sheet. Moreover, the activity proved to be useful to introduce more complex topics related to GR, such as the phenomenon of gravitational lensing and black holes. Another important aspect is that the rubber sheet makes it easier to connect apparently distant phenomena, such as the motion of the planets and the way a black hole looks, thus reinforcing the idea that the new Einsteinian description is perfectly coherent with the classical and most familiar framework of gravity (an idea that is very far from the common way of thinking, even that of teachers sometimes).
These considerations help us to select the topics that could have been effectively treated with teachers. In particular, we selected Kepler's laws, gravity assist, binary systems, gravitational lensing and black holes. We also decided to dedicate some time to provide teachers with further ideas on the use of the rubber sheet not explored in depth, such as the formation of the Solar System, time dilation and gravitational waves.
Another important aspect that we had to consider when using the RSA with teachers was the possible birth of misconceptions and wrong beliefs related to gravity. It is known from the literature \cite{Freefall, Curved, Price, Gould}, for example, that one problematic aspect of the rubber sheet consists in leading to think that space-time is two-dimensional, and that gravity always points downward, as happens on Earth. In fact, the rubber sheet is almost two-dimensional, and the marble thrown on it orbits around the weight placed in the centre and finally falls toward it. The same word ``black holes" suggest an object, a hole, into which matter falls. ``My marble can escape the gravitational attraction of the black hole: I can throw it upwards so that it leaps over the hole" is a thought that participants often shared with us during the experimentation with high school students. Of course, gravity doesn’t work like that: matter is simply attracted in space, there is no bottom or top, above or below, but rather near/far in all directions. Black holes are spherical objects that attract the surrounding matter in all directions. Unfortunately, there is no way to visualize this using the rubber sheet. But this limitation can be made explicit, and students' considerations (as the one cited above) can be used to make them reflect. This is the approach we used, and which proved to be effective: only 3\% of our sample chose the distractor ('a force that points downward') between the alternatives when asked to specify what deforms space-time in the questionnaires administered after the activity \cite{Postiglione1}. We thus understood that we could suggest this strategy to the teachers.
Another critical point of the RSA is the fact that GR foresees the deformation of space and time, while the rubber sheet allows to only visualize the spatial deformation \cite{Curved}. Similarly, to the misconception analysed above, this is an intrinsic limitation of the model that cannot be overcome. One way to deal with it, however, is to explicitly address it and raise a discussion. 
These reflections led us to understand the importance of dedicating a relevant part of our work with the teachers to the definition of the model used, the simplifications introduced and to the way to guide the discussion with students so that they can really understand the topics treated without giving birth to wrong beliefs.
Once the issues to be covered were clear, we used another teachers’ training course of the Department to test, optimize and finalize the methodology we developed. In particular, after a long and productive debate, we defined the guidelines and tricks necessary for the teacher to autonomously carry out our activity with the rubber sheet in their classrooms. As a last step, we thought of a structure for the space-time simulator that could be cheap, easy to build, and that could fit comfortably in the classroom. 
At the end of our work, all the material was collected in a manual that was published in the form of an open access eBook \cite{Postiglione_book}. The manual, designed at first for the Italian school, is written in Italian but also in English, in order to reach a wider audience, and is divided in 8 educational cards, each of which deals with a certain topic. The first card provides the instructions to assemble the space-time structure, using a hula-hoop, some wooden strips and a lycra sheet. Then, the concept of space-time and the model of the RSA are introduced, together with all the simplification implied and the suggestions that can help the teacher guide students to notice and overcome misconceptions. The other cards deal with specific topics related to Newtonian gravity (Kepler's laws, binary systems, gravity assist) and GR (gravitational lensing, black holes, time dilation). All the topics covered are treated by guiding the teacher step by step, starting from the materials to be used to show the phenomenon, how to use them, how to interact with students, and how to provide more insights making use of appropriately chosen photos and videos. A more complete description of the manual can be found in a dedicated paper \cite{Postiglione2}.
In order to evaluate the response to the manual by readers, we added to it two questionnaires. The first one was designed to be completed after reading the book and concerns the readers’ evaluation on the clarity and completeness of the description of the activities proposed, in addition to probing the teacher's intention to carry out the activities. The second questionnaire is instead dedicated to the readers who have already carried out the activity in the classroom and who can therefore evaluate the feasibility of realization, the students’ reaction and the efficacy in explaining the physics concepts.

\section{First reactions to the developed material}
The manual we developed is available online and until April 2021, more than 300 copies have been downloaded. In April 2021 we had received 22 responses to the first questionnaire. From these answers we can gain a first idea of the teachers’ reactions to the material we have developed.
In particular, the readers think that the book is interesting (the 68\% gives a score of 5/5, the remaining gives a score of 4/5), that the activities are presented clearly and in a complete way (72\% gives 5/5, the remaining 4/5). Moreover, the cards are useful (72\% gives 5/5, 23\% gives 4/5 and 4\% gives 3/5) and the content of the cards seem sufficient to carry out the activities in your classroom (68\% gives 5/5, 27\% gives 4/5). Regarding the simplicity of construction of the structure, the 40\% gives a score of 5/5, the 60\% of 4/5 while the remaining gives a score 3/5 (4\%) or 2/5 (4\%).
The 66\% of the readers plans to realize the activity in the classroom (with a 23\% of “definitely yes” and the remaining not sure about it), especially the ones dealing with the general introduction to the space-time, Kepler's laws (42,9\%), gravity assist (33\%), black holes (28\%). However, the 38\% shares the intention of trying all the activities proposed. The reasons are many: the proposal is interesting and easy to realize, fun, useful, in line with the school program. 
Regarding the year during which carrying out the activities, the 61\% propose the fifth and last year of high school, as an introduction to Modern Physics, but also the third year (52\%) when treating Kepler's laws.
Among the topics treated, Kepler's laws and black holes are the most cited among the favourites, while binary systems, gravitational lensing and time dilation seem less preferred and therefore can be considered only as further insights.
These comments are in line with the considerations that came out from a focus group of teachers we organized to talk about the manual. In particular, these teachers appreciated the structure of the book, the way the activities are described, and claim to be very curious to try the activities with their classes as soon as possible to give a more precise opinion.
Although the answers to the first questionnaire show the teachers’ will of experimenting the activity, we have not received yet answers to the second questionnaire, related to the realization of the activities by the teachers. This is probably also due to the current difficulty of carrying out activities in presence in schools due to the Covid-19 emergency.

\section{Conclusions}
In this paper we presented a teachers’ professional development approach we realized at the Department of Mathematics and Physics of Roma Tre University with the aim of helping high school teachers to treat gravity in their classes using the rubber sheet analogy. Our methodology then resulted in a manual that contains the materials developed, that is freely available online.Unfortunately, the Covid-19 emergency prevented us from following the teachers in their experimentation with their students. We believe, however, that our work represents a significant contribution to the demand of treating gravity using its most modern approach, General Relativity, and thus of introducing Modern Physics in high school. In particular, we believe that our proposal can help teachers to be autonomous in dealing with the topics proposed using interactive and engaging activities. 
Our approach to gravity using the RSA proved to be valid and effective when tested with high school students, because it helps them to visualize, comprehend and remember classical Newtonian gravity as well as some of the topics foreseen by GR. Regarding teachers, we believe that our approach was useful because it allowed a continuous debate with them on this topic; during the numerous discussions we had, teachers always declared to be willing to carry on the proposed activities with their students following the manual we developed, and this is also confirmed by the first results of the questionnaire attached to the book. Among the topics proposed, teachers seem to prefer Kepler's law and black holes, indicating that the rubber sheet can be used both to visualize and reinforce topics already typically treated in school, and to introduce new and more complex topics.
For the future, we plan to continue our training with other teachers also in the courses we will organize in presence. Furthermore, we want to closely follow the activity as carried out by the teachers who already collaborated with the Department, investigating the clarity and effectiveness of the material we have collected in the manual. Finally, we plan to continue to monitor the evaluations of the activities as provided by the readers of the book.

\end{document}